\documentclass[]{article}
\voffset= - 1.3in
\hoffset= - 1.2in         

\makeatletter
\newdimen\normalarrayskip              
\newdimen\minarrayskip                 
\normalarrayskip\baselineskip
\minarrayskip\jot
\newif\ifold             \oldtrue            \def\new{\oldfalse}
\def\arraymode{\ifold\relax\else\displaystyle\fi} 
\def\eqnumphantom{\phantom{(\theequation)}}     
\def\@arrayskip{\ifold\baselineskip\z@\lineskip\z@
     \else
     \baselineskip\minarrayskip\lineskip2\minarrayskip\fi}
\def\@arrayclassz{\ifcase \@lastchclass \@acolampacol \or
\@ampacol \or \or \or \@addamp \or
   \@acolampacol \or \@firstampfalse \@acol \fi
\edef\@preamble{\@preamble
  \ifcase \@chnum
     \hfil$\relax\arraymode\@sharp$\hfil
     \or $\relax\arraymode\@sharp$\hfil
     \or \hfil$\relax\arraymode\@sharp$\fi}}
\def\@array[#1]#2{\setbox\@arstrutbox=\hbox{\vrule
     height\arraystretch \ht\strutbox
     depth\arraystretch \dp\strutbox
     width\z@}\@mkpream{#2}\edef\@preamble{\halign
\noexpand\@halignto
\bgroup \tabskip\z@ \@arstrut \@preamble \tabskip\z@ \cr}%
\let\@startpbox\@@startpbox \let\@endpbox\@@endpbox
  \if #1t\vtop \else \if#1b\vbox \else \vcenter \fi\fi
  \bgroup \let\par\relax
  \let\@sharp##\let\protect\relax
  \@arrayskip\@preamble}
\def\eqnarray{\stepcounter{equation}%
              \let\@currentlabel=\theequation
              \global\@eqnswtrue
              \global\@eqcnt\z@
              \tabskip\@centering
              \let\\=\@eqncr
 \halign to \displaywidth\bgroup
    \eqnumphantom\@eqnsel\hskip\@centering
    $\displaystyle \tabskip\z@ {##}$%
    \global\@eqcnt\@ne \hskip 2\arraycolsep
         $\displaystyle\arraymode{##}$\hfil
    \global\@eqcnt\tw@ \hskip 2\arraycolsep
         $\displaystyle\tabskip\z@{##}$\hfil
         \tabskip\@centering
    &{##}\tabskip\z@\cr}
\begingroup\ifx\undefined\newsymbol \else\def\input#1 {\endgroup}\fi
\newfont{\hr}{msbm10}
\newfont{\ams}{msam10}

\font\numbers=cmss12
\font\upright=cmu10 scaled\magstep1
\def\stroke{\vrule height8pt width0.4pt depth-0.1pt}
\def\topfleck{\vrule height8pt width0.5pt depth-5.9pt}
\def\botfleck{\vrule height2pt width0.5pt depth0.1pt}
\def\Zmath{\vcenter{\hbox{\numbers\rlap{\rlap{Z}\kern 0.8pt\topfleck}\kern
2.2pt
                   \rlap Z\kern 6pt\botfleck\kern 1pt}}}
\def\Qmath{\vcenter{\hbox{\upright\rlap{\rlap{Q}\kern
                   3.8pt\stroke}\phantom{Q}}}}
\def\Nmath{\vcenter{\hbox{\upright\rlap{I}\kern 1.7pt N}}}
\def\Cmath{\vcenter{\hbox{\upright\rlap{\rlap{C}\kern
                   3.8pt\stroke}\phantom{C}}}}
\def\Rmath{\vcenter{\hbox{\upright\rlap{I}\kern 1.7pt R}}}
\def\Z{\ifmmode\Zmath\else$\Zmath$\fi}
\def\Q{\ifmmode\Qmath\else$\Qmath$\fi}
\def\N{\ifmmode\Nmath\else$\Nmath$\fi}
\def\C{\ifmmode\Cmath\else$\Cmath$\fi}
\def\R{\ifmmode\Rmath\else$\Rmath$\fi}

\def\stackreb#1#2{\mathrel{\mathop{#2}\limits_{#1}}}
\def\Tr{{\rm Tr}}
\def\res{{\rm res}}
\def\Bf#1{\mbox{\boldmath $#1$}}

\def\bbeta{{\Bf\beta}}

\def\bPhi{{\Bf\Phi}}

\def\bxi{{\Bf\xi}}
\def\bbeta{{\Bf\eta}}
\def\d{\partial}

\def\Im{{\rm Im}}

\def\diag{{\rm diag}}
\def\2{{1\over 2}}
\def\N2{${\cal N}=2$}
\def\4N{${\cal N}=4$}
\def\1N{${\cal N}=1$}

\def\be{ \begin{eqnarray} }
\def\ee{ \end{eqnarray} }
\def\d{\partial}
\def\bea{\begin{eqnarray}}
\def\eea{\end{eqnarray}}

\def\beq{\begin{equation}}
\def\eeq{\end{equation}}
\def\ba{\beq\new\begin{array}{c}}
\def\ea{\end{array}\eeq}
\def\be{\ba}
\def\ee{\ea}
\def\stackreb#1#2{\mathrel{\mathop{#2}\limits_{#1}}}

\begin{document}

\begin{flushright}
FIAN/TD-05/99\\
ITEP/TH-23/99
\end{flushright}
\vspace{0.5cm}
\begin{center}
{\LARGE\bf SUSY gauge theories and Whitham integrable systems
         from compactification and SUSY breaking
\footnote{Contribution to the proceedings of the workshop
          "Gauge Theory and Integrable Systems", YITP, Kyoto,
          26-29 January, 1999}
}\\
\vspace{0.5cm}
{\Large
A.Marshakov
\footnote{e-mail address: mars@lpi.ac.ru,\
andrei@heron.itep.ru}}\\
\vspace{0.5cm}
{\it Theory Department, Lebedev Physics Institute, Moscow ~117924, Russia
\\ and \\ ITEP, Moscow 117259, Russia},
\\
\end{center}
\bigskip
\begin{quotation}
We review the Seiberg-Witten construction of low-energy effective actions
and BPS spectra in SUSY gauge theories and its formulation in terms of
integrable systems. It is also demonstrated how this formulation naturally
appears from the compactified version of the theory with partially broken
supersymmetry so that the integrable structures arise from the relation
between bare and quantum variables and superpotentials of SUSY gauge
theories. The Whitham integrable systems, literally corresponding to the
uncompactified theory, are then restored by averaging over fast variables in
the decompactification limit.
\end{quotation}

\setcounter{footnote}{0}
\setcounter{equation}{0}
\section{Introduction}

Several past years brought us new understanding of non-perturbative
phenomena in supersymmetric (SUSY) quantum gauge theories. In particular it
has become possible to take into account all instanton effects and write down
the {\em exact} low-energy effective actions in \N2 SUSY Yang-Mills theories
\cite{SW1,SW2}. The proposed non-perturbative formulas imply an existence of
underlying hidden geometric structures and, in a most elegant way, can be
formulated in terms of integrable systems \cite{GKMMM}.
This question has already a long story, but the {\em origin} of this
relation still remains to be an open problem. The aim of these notes is, in
particular,
to discuss and partially fill this gap. We shall see, for example, that
some aspects of this relation can be
more clearly understood if one takes, first, SUSY gauge theory in the
space-time with compactified dimensions \cite{SW3}.

The reason is that the compactified gauge theory has larger moduli space than
its fully non-compact relative \cite{SW3}, and this moduli space can be
thought of as a phase space of certain classical integrable system. We shall
consider the compactified theory with broken SUSY by the Scherk-Schwarz
mechanism (down to \1N in 4D sense), i.e. when the breaking mass parameter is
given by ${\epsilon\over R}$, where $R$ is the radius of compact dimension
and $\epsilon$ -- phase parameter of boundary conditions. In the
decompactification limit $R\to\infty$ supersymmetry is restored but the
extra, compact, degrees of freedom become "heavy" and the integration over
them leads to the "averaging" of an integrable system in the
Bogolyubov-Whitham sense -- thus becoming an origin of the relation
between Seiberg-Witten theories and Whitham integrable systems \cite{GKMMM}.

\section{Integrable systems in Seiberg-Witten theory
\label{s:swis}}

We start with the
{\em formulation} of the exact effective actions for the 4D SUSY gauge
theories {\em a la} Seiberg and Witten (SW) \cite{SW1,SW2} which is very
simple:  the (Coulomb branch) low-energy effective action for the \N2 SUSY
Yang-Mills vector multiplets (supersymmetry requires the metric on moduli
space of massless complex scalars from \N2 vector supermultiplets to be of a
"special K\"ahler form" -- or the K\"ahler potential $K({\bf a},{\bar{\bf
a}}) = \Im\sum_i {\bar a}_i {\d{\cal F}\over\d a_i}$ should be expressed
through a {\em holomorphic} function ${\cal F} = {\cal F}(\bf a)$ -- a {\em
prepotential}) can be described in terms of auxiliary Riemann surface
(complex curve) $\Sigma $, endowed with a meromorphic 1-differential $dS$,
which possess peculiar properties:

\begin{itemize}

\item The number of "live" moduli (of complex structure) of $\Sigma $ is
strongly restricted (roughly "3 times" less than for generic Riemann surface).
The genus of $\Sigma $ -- for the $SU(N)$ gauge theories -- is exactly
equal
to the rank of gauge group -- i.e. to the number of independent moduli.

\item The variation of generating 1-form $dS$ over these moduli gives
{\em holomorphic} differentials
\be
\delta_{\rm moduli}dS = {\rm holomorphic}
\label{hol}
\ee

\item The (canonical) ${\bf A}$- and ${\bf B}$-periods of generating 1-form
\be
{\bf a} = \oint_{\bf A}dS
\ \ \ \ \ \
{\bf a}_D = \oint_{\bf B}dS
\label{periods}
\ee
give the set of "dual" BPS masses -- the W-bosons and the monopoles while the
period matrix $T_{ij}(\Sigma)$ -- the set of couplings in the low-energy
effective theory. From (\ref{hol}) and (\ref{periods}) one gets the relation
between the BPS masses and couplings
\be
{\d a_D^i\over\d a_j} = \oint_{B_i}d\omega_j = T_{ij}(\Sigma)
\ee

\item All above requirements can be summarized saying that the SW data
are equivalent to defining an {\em integrable system} \cite{GKMMM} in the
sense of \cite{DKN}. The periods (\ref{periods}) are the action variables
and the (holomorphic) variation of the generating 1-form (\ref{hol}) gives
rise to the dual (angle) variables.
The corresponding class of integrable models include well-known integrable
systems of particles (the periodic Toda chains, Calogero-Moser models and
their relativistic Ruijsenaars generalizations) and classical spin chains
(see \cite{SWRev,M} and references therein for details).

\item The prepotential is function of half of the variables
(\ref{periods}), say ${\cal F} = {\cal F}({\bf a})$, then
\be
a_D^i = {\d{\cal F}\over\d a_i}
\\
T_{ij} = {\d a_D^i\over\d a_j} = {\d^2{\cal F}\over\d a_i\d a_j}
\label{adt}
\ee
The prepotential ${\cal F}$ itself has {\em no natural} definition in the
language of classical finite-gap integrable systems \cite{DKN} -- in order
to describe it one has to consider {\em deformations} of the finite-gap
models. It is possible to show, for example, that it satisfies the
following system of differential equations \cite{MMM2}
\be
{\cal F}_i {\cal F}_k^{-1} {\cal F}_j = {\cal F}_j {\cal F}_k^{-1} {\cal F}_i
\ \ \ \ \ \ \forall i,j,k = 1,\ldots,N-1.
\label{FFF}
\ee
where ${\cal F}_i$ denotes the matrix of the third derivatives
\be
({\cal F}_i)_{mn} = \frac{\partial^3 {\cal F}}{\partial a_i
\partial a_m\partial a_n} = {\d T_{mn}\over\d a_i}
\ee
or
\be
{\cal F}_i G^{-1} {\cal F}_j = {\cal F}_j G^{-1} {\cal F}_i
\ \ \ \ \ \ \forall i,j,k = 1,\ldots,N-1.
\label{FFFG}
\ee
where $G = \sum_k g_k{\cal F}_k$ for any $g_k$. The system of equations
(\ref{FFF}) holds {\em non-perturbatively}
for most of SW prepotentials (an important exception is the case of broken
\4N theory -- the elliptic Calogero-Moser model \cite{WDVV-long}). The proof
of the Eqs.~(\ref{FFF}), (\ref{FFFG}) is based on the existence of closed
algebra of multiplication of holomorphic differentials on certain Riemann
surfaces \cite{MMM2,WDVV-long}, which is a sort of generalization of
polynomial rings.

\item The SW prepotential ${\cal F}$ can be also considered as a particular
case of the tau-function of the Whitham hierarchy \cite{KriW} -- a function of
infinitely many extra parameters $T_n$ so that in the "SW point"
\be
{\cal F}_{SW}({\bf a},\Lambda) \equiv \left.{\cal F}({\bf a},{\bf
T})\right|_{T_n=\Lambda\delta_{n,1}}
\ee
It implies, in particular, that the generalized prepotential satisfies
\cite{rgwhi}
\be
\frac{\partial{\cal F}}{\partial T_n} = \frac{N}{2\pi i n}
\sum_m  mT_m {\cal A}_{mn}
= \frac{N}{2\pi in}T_1{\cal H}_{n+1} + O(T_2,T_3,\ldots)
\\
\frac{\partial^2{\cal F}}{\partial a_i\partial T_n}
= \frac{N}{2\pi in} \frac{\partial {\cal H}_{n+1}}{\partial a_i}
\\
\frac{\partial^2{\cal F}}{\partial T_m\partial T_n}
= -\frac{N}{2\pi i} \left({\cal A}_{mn}
+ \frac{N}{mn}\frac{\partial {\cal H}_{m+1}}{\partial a_i}
\frac{\partial {\cal H}_{n+1}}{\partial a_j}
\partial^2_{ij} \log \theta_E(0| T)\right)
\label{2der}
\ee
where
\be
{\cal A}_{mn} = -\frac{N}{mn}
{\res}_\infty\left(P^{n/N}(\lambda)d P^{m/N}_+(\lambda)\right)
= {\cal A}_{nm}
\ee
and
\be
{\cal H}_{n+1} \equiv {\cal A}_{n1}
= -\frac{N}{n}{\res}_\infty P^{n/N}(\lambda)
d\lambda  = h_{n+1} + O(h^2)
\ee
The most illustrative form is
\be
\frac{\partial^2}{\partial T_m\partial T_n}
\left({\cal F}(a,T) -
\frac{N}{4\pi i}{\cal F}^{GKM}(a,T)\right)=
\\
= -\frac{N^2}{2\pi imn}
\frac{\partial {\cal H}_{m+1}}{\partial a_i}
\frac{\partial {\cal H}_{n+1}}{\partial a_j}
\partial^2_{ij} \log \theta_E(0| T)
\label{2dergkm}
\ee
where
\be
{\cal F}^{GKM}(a|T) \equiv \frac{1}{2} \sum_{m,n} T_mT_n
{\cal A}_{mn}
\ee
is the prepotential of Generalized Kontsevich Model \cite{LGGKM} --
a 2D topological string theory or the "local" part which is not related to
the structure of nontrivial SW spectral curve, while the r.h.s. of
(\ref{2dergkm}) is expressed through the derivatives of $\theta$-constant,
corresponding to a particular SW curve (and certain choice of characteristic
on this particular curve).

\item Eqs.~(\ref{FFF}), (\ref{FFFG}), (\ref{2der}) and (\ref{2dergkm}) are
{\em classical} differential equations one can write for the {\em quantum}
effective actions which, in particular, depend on the Planck constant
$\hbar$ or the string scale $\sqrt{\alpha'}$. The same phenomenon was
studied earlier in the case of 2D topological string models (see,
for example \cite{2D} and references therein).

\end{itemize}

These properties of the low-energy effective SUSY gauge theories follow
from the Seiberg-Witten hypothesis and were derived using calculus on
Riemann surfases. However they do not tell us anything about the {\em
origin} of this non-perturbative structure. In particular the physical sense
of arising integrable system remains to be unclear. One of the ways to try
to understand it is to consider the compactified version of the SW theory.

There were two, at first glance different, consequences of adding
a compact dimension in the context of Seiberg and Witten. A
straightforward one is to add 5-th compact dimension to 4D SUSY gauge theory
and to take into account
the contribution of the soft Kaluza-Klein (KK) modes \cite{Ne96} which leads to the
"relativization" of an integrable system. A different effect of enlarging
moduli space arises when one considers (preserving SUSY!) compactification
down to 3+1 (compact) dimension \cite{SW3}. Both ways are particular
cases of generic compactified theory with Wilson lines, or new moduli -- the
monodromies of gauge fields and, in general, such compactification breaks
supersymmetry (at least partially), giving rise, in particular, to new
effective theories of SW type \cite{BMMM1,BMMM2,M99}.

In order to discuss compactified theory, we shall consider, first, general
properties of moduli spaces (vacua) of SUSY
Yang-Mills, or, if speaking about bosonic sector, the so called
Yang-Mills-Higgs theories. On these moduli spaces one can introduce
holomorphic symplectic 2-forms which allow to construct an integrable
system in the sense of \cite{DKN} or \cite{Hi}. The integrating change
of variables can be considered as relation between {\em bare} or classical and
{\em quantum} variables in the context of corresponding gauge theory.
Partially this relation is known as a relation between "bare" moduli
$u_k = {1\over k}\Tr\Phi^k$ and the "exact" quantum moduli -- the periods of
Seiberg-Witten differential or action variables in corresponding integrable
system. The rest part of this relation --
exactly the integrating change of variables in finite-gap integrable system
has also the meaning of relation between the "bare" Wilson loops (or dual
moduli) and their exact quantum counterparts. We shall also discuss the
symmetry between Wilson loops and scalar moduli under T-duality
transformation and show that it is related with the duality between the
co-ordinates and action variables in integrable systems
\cite{Fock,GNR}.

As a particular example of compactified theory we shall consider 3+1
dimensional vector supermultiplet with Wilson line softly breaking SUSY
from \N2 in 4D sense down to \N2 in 3 dimensions (from two complex Weyl
spinor supercharges to one complex or two real Weyl supercharges). This,
macroscopically 3-dimensional theory is known to generate Toda chain
superpotential \cite{Pol,Wad,AHW} and we shall
demonstrate that this superpotential is directly related to the Toda chain
dynamics arising in the context of SW theory.

\section{Perturbative gauge theories and "degenerate" integrable systems
\label{s:pert}}

The relation between SW theories and integrable systems can be already
discussed at the {\em perturbative} level \cite{M99}, where \N2 SUSY
effective actions are completely defined by the 1-loop contributions (see
\cite{SW1} and references therein). The scalar field $\bPhi = \|\Phi_{ij}\|$
of \N2 vector supermultiplet acquires nonzero VEV $\bPhi = {\rm
diag}(\phi_1,\dots,\phi_N)$ and the masses of ``particles'' --
$W$-bosons and their superpartners are proportional to
$\phi_{ij}=\phi_i-\phi_j$ due to
the Higgs term $[A_\mu,\bPhi]_{ij} = A_\mu^{ij}(\phi_i-\phi_j)$ in the SUSY
Yang-Mills action. These masses can be written altogether in terms of the
generating polynomial
\be
w = P_N(\lambda ) = \det (\lambda - \bPhi) = \prod (\lambda - \phi_i)
\label{polyn}
\ee
where $\bPhi $ is the adjoint complex scalar ($\Tr\bPhi = \sum_i\phi_i = 0$),
via residue formula
\be\label{pertsample}
m_{ij} \sim \oint_{C_{ij}} \lambda d\log w=
\oint_{C_{ij}} \lambda d\log P_{N}(\lambda )
\ee
which for a particular "$\infty$-like" contour $C_{ij}$ around the roots
$\lambda = \phi_i$ and $\lambda = \phi_j$ gives rise exactly to the Higgs
masses.  The contour integral (\ref{pertsample}) is defined on a complex
$\lambda $-plane with $N$ removed points: the roots of the polynomial
(\ref{polyn}) -- a {\em degenerate} Riemann surface. The masses of monopoles
are naively infinite in this limit, since the corresponding contours (dual to
$C_{ij}$) start and end in the points where $dS$ obeys pole singularities. It
means that the monopole masses, proportional to the squared inverse coupling,
are renormalized in perturbation theory and defined naively up to the masses
of particle states times some divergent constants.

The effective action (the prepotential) ${\cal F}$, or the set of effective
charges $T_{ij}$ (\ref{adt}), are defined in \N2 perturbation theory
completely by 1-loop diagram giving rise to the logarithmic corrections
\be\label{effcharge}
\left(\delta ^2{\cal F}\right)_{ij} = T_{ij}
 \sim \sum_{\rm masses} \log {({\rm mass})^2\over\Lambda ^2} =
\log {(\phi_i-\phi_j)^2\over\Lambda ^2}
\ee
where $\Lambda\equiv\Lambda_{QCD}$ and last equality is true only for pure
gauge theories -- since the only masses we have there are given by
(\ref{pertsample}). That is all one has
in the perturbative {\em weak-coupling} limit of the SW construction, when
the instanton contributions to the prepotential (being proportional to
the degrees of  $\Lambda^{2N}$ (or $q^{2N}\equiv e^{2\pi i\tau N}$ -- in the
UV-finite theories with bare coupling $\tau $)
are (exponentially) suppressed so that one keeps only the terms proportional
to $\tau $ or $\log\Lambda$. These degenerated {\em rational} spectral curves
can be already related to the family of {\em trigonometric}
Ruijsenaars-Schneider and Calogero-Moser-Sutherland systems and the {\em
open} Toda chain or Toda molecule.

For example, in the case of $SU(2)$ pure gauge theory Eq.~(\ref{polyn})
turns into
\be
w = \lambda^2 - u
\label{polsu2}
\ee
with $u = \2\Tr\bPhi^2$. In the parameterization of \cite{SW1}
$X = w = e^z = \lambda^2 - u$, $Y = w\lambda$ the same equation can be
written as
\be
Y^2 = X^2(X+u)
\label{optoda}
\ee
and the masses (\ref{pertsample}) are now defined by the contour integrals of
\be
dS = \lambda d\log w = 2{\lambda^2d\lambda\over\lambda^2-u}
= {\lambda d\lambda\over\lambda - \sqrt{u}} +
{\lambda d\lambda\over\lambda + \sqrt{u}}  = \sqrt{X+u}~{dX\over X}
\label{dSop}
\ee
One can easily notice that Eqs.~(\ref{polsu2}), (\ref{optoda}) and
(\ref{dSop}) can be interpreted as {\em integration} of the open
$SL(2)$ (the Liouville) Toda chain with the co-ordinate $X=w=e^q$, momentum
$p=\lambda$ and Hamiltonian (energy) $u$. The integration of generating
differential $dS = pdq$ over the trajectories of the particles gives rise,
in fact, to the monopole masses in the SW theory.

This is actually a general rule -- the perturbative \N2 theories of the
"SW family" give rise to the "open" or trigonometric family of integrable
systems -- the open Toda chain, the trigonometric Calogero-Moser or
Ruijsenaars-Schneider systems. This can be easily established
at the level of spectrum (\ref{pertsample}) and the effective couplings
(\ref{effcharge}) -- the corresponding (rational) curves are (\ref{polyn})
in the $N$-particle Toda chain case
\be
w = {P^{(CM)}_N(\lambda )\over P^{(CM)}_N(\lambda + m)}
\ \ \ \ \ \ \ \ \
dS = \lambda{dw\over w}
\label{triCa}
\ee
for the trigonometric Calogero-Moser-Sutherland model and
\be
w=\frac{P^{(RS)}_N(\lambda)}
{P^{(RS)}_N(\lambda e^{-2i\epsilon})}
\ \ \ \ \ \ \ \ \
dS = \log\lambda {dw\over w}
\label{triRu}
\ee
for the trigonometric Ruijsenaars-Schneider system. It is easy to see that
(perturbative) spectra are given by general formula \cite{BMMM2}
\be
M = \phi_{ij} \oplus {\pi n\over R} \oplus {\epsilon +\pi n\over R}
\\
N\in {\bf Z}.
\label{spectrum}
\ee
and contain in addition to the Higgs part $\phi_{ij}$ the KK
modes $\pi n\over R$ and the KK modes for the fields with "shifted" by
$\epsilon$ boundary conditions. The $\epsilon$ parameter can be treated as a
Wilson loop of gauge field along the compact dimension and in a subclass of
models \cite{SWCal} ${\epsilon\over R}$ plays the role of the mass of the
adjoint matter multiplet.

Thus we see that the relation between effective actions of SUSY gauge
theories and classical integrable systems is  really established on {\em
perturbative} level. Moreover, the perturbative limit can be considered as
a self-consistent approximation since all ingredients of the relation
between the SW theories and integrable systems we mentioned above can be
consistently (and even {\em explicitly}), in particular:

\begin{itemize}

\item The associativity equations (\ref{FFF}) possess an obvious class of {\em
perturbative} solutions, for example (cf. with (\ref{effcharge}))
\be
{\cal F}_{\rm pert} = \frac{1}{2}\sum_{\stackrel{i<j}{i,j=1}}^{N-1}
(a_i-a_j)^2\log(a_i-a_j) +
\frac{1}{2}\sum_{i=1}^{N-1}a_i^2\log a_i
\label{pertF}
\ee
(see a proof by straightforward calculation in \cite{MMM2}), which is a direct
weak-coupling limit of the full SW prepotential when the non-perturbative
terms (powers of $\Lambda$) are suppressed. A generic approach to the
computation of perturbative prepotentials (including the theories with
the KK excitations) based on SW theory can be found in \cite{MaMi97}.

\item Whitham hierarchy has a particular class of solutions corresponding to
degenerate Riemann surfaces $\Sigma$ where handles turn into (pairs of)
points and holomorphic differentials into the differentials with first-order
poles at these points. The generating differential in this case has general
structure
\be
dS = \sum T_n\lambda^{n-1}d\lambda + \sum
a_i{d\lambda\over\lambda-\lambda_i}
\ee
This formula provides a straightforward way of computation of the
"perturbative" Whitham tau-function with non-zero times $T_n$
\footnote{However, it would be nice to have an explicit expression.}.

\end{itemize}

Despite one can really check in the perturbative limit all the
statements of sect.~\ref{s:swis} by standard field theory computation, the
nature of this relation at such level remains unclear.

\section{String theory and Yang-Mills-Higgs system}

In general,
moduli spaces of SUSY gauge theories are described in terms of the
(eigenvalues of the) scalar fields from SUSY multiplets and  -- in general
situation when space-time has compact dimensions -- one should also add
the Wilson loops of gauge fields themselves. The scalar Higgs fields in
string picture are associated with the positions of branes (the
hypersurfaces) in some transverse directions and the effective world-volume
gauge theories
are coming from open strings ending on D-branes. The quantum moduli spaces of
branes are formulated in terms of {\em full} matrices (rather than their
eigenvalues) \cite{WittD} with the potential
\footnote{Moduli space can be also described in terms of holomorphic
{\em superpotential}, for example, for 6 real fields
\be
W(\Phi ) = \Tr\epsilon_{ijk}{\tilde\Phi}_i [{\tilde\Phi}_j,
{\tilde\Phi}_k]
\label{supot}
\ee
where $\tilde\Phi_i = \Phi_{i_1}+i\Phi_{i_2}$ are already complex scalars.}
\be
V(\Phi ) = \Tr\sum_{i<j} [\Phi_i,\Phi_j]^2
\label{pot}
\ee
which arises in the action under the compactification of 10D \1N SUSY
Yang-Mills theory
\be
\Tr\int_{d^{10}x}{\bf F}_{MN}^2 + {\rm fermions}
\ee
down to 4D (\4N SUSY) theory
\be
\Tr\int_{d^4x}{\bf F}_{\mu\nu}^2 + (D_\mu\Phi_i)^2 +
\sum_{i<j}[\Phi_i,\Phi_j]^2 + {\rm fermions}
\label{ymh}
\ee
so that $M=(\mu,i)$, i.e. $A_\mu = A_M$ for $M=0,\dots,3$ and $\Phi_i = A_M$
for $M=4,\dots,9$.
In the vacuum sector one can neglect the fermionic terms in (\ref{ymh}) and,
thus, speak about the {\em Yang-Mills-Higgs} system. Minima of the potential
(\ref{pot}) correspond to $[\Phi_i,\Phi_j]=0$, or to the simultaneously
diagonal matrices
\be
\Phi_i = \diag~(\phi^{(i)}_1,\ldots,\phi^{(i)}_N)
\label{diagma}
\ee
Distinct eigenvalues correspond to $U(1)^{N-1}$ gauge theory, when the
eigenvalues of $\Phi_i$ coincide the gauge symmetry is restored up to
$SU(N)$. In terms of modern string theory, the action (\ref{ymh}) literally
corresponds to a system of $N$ parallel D3-branes or to (bosonic sector of)
\4N SUSY Yang-Mills theory. Six scalars $\Phi_i$ can all obey non-zero VEV's
so that the dimension of moduli space is $6(N-1)$. D3 branes can be thought
of as D5 branes $X_0,\dots,X_5$ compactified on two-torus, if, say, the
dimensions $X_4$ and $X_5$ are compact.

In theories with \4N SUSY in four dimensions effective couplings and BPS
masses are not renormalized -- the theory has lots of symmetries including
conformal group and there are no dimensionful (mass) parameters. It means,
in particular that the eigenvalues of the scalar fields $\{\phi_i\}$ are
not renormalized, as well as $\{ q_i\}$ -- the eigenvalues of the Wilson
loops $\oint A_\mu dx^\mu$ -- if we consider a theory with compact dimensions.
The correspondence between SW theories and integrable systems implies that
theory with \4N SUSY corresponds to a system of {\em free} particles -- with
the Hamiltonians $u_k = {1\over k}\Tr\Phi^k = {1\over k}\sum_i\phi_i^k$,
where $\{\phi_i\}$ play the role of momenta. If one adds compact dimensions,
the extra moduli $\{ q_i\}$ can play the role of dual co-ordinates -- at
least in the sense that the volume of moduli space is (non-renormalized,
dimensionless) constant and the volume form can be presented as a degree
of symplectic form
\be
\Omega = \Tr\int\delta{A}\wedge\delta\Phi = d\phi_i\wedge dq_i
\label{bss}
\ee
Breaking
supersymmetry down to \N2 the potential acquires extra mass terms
$m_i^2\Tr\Phi_i^2$ for two of three scalar fields and the dimension of
"scalar" moduli space goes down to $2(N-1)$ or to one complex "diagonal
matrix" field (\ref{diagma}). Moreover, in contrast to \4N theory, in general
case (of nontrivial boundary conditions) matrices of scalar fields and
monodromies become dependent of each other, or satisfy nontrivial
commutation relation like
\be
[A,\Phi] \sim mJ
\label{momap}
\ee
linear in the parameter of "massive deformation" \cite{SWCal} (for $m\to 0$
one comes back to \4N theory) and $J$ is some matrix of
"gauge-covariant" form.

Quantum effects turn the bare symplectic form (\ref{bss}) into
\be
\sum \delta q_i\wedge\delta p_i \mapsto \sum\delta \vartheta_k\wedge\delta
a_k
\label{qss}
\ee
where
\be
a_i = \oint_{A_i}dS
\ee
are correct quantum variables -- the Seiberg-Witten integrals.
One may consider $a_i = a_i(\Phi,\Lambda)$
as a transformation from bare quantities -- the eigenvalues $\{\phi_i\} $ to
their exact quantum values in the effective theories $\{a_i\}$, playing the
role of the exact quantum BPS masses. In the same way one should consider the
transformation $q_i\rightarrow\vartheta_i$ as transformation from bare
value of monodromy to its exact quantum value in the effective theory.

\section{Compactification to 3+1 dimensions and SUSY breaking}

SUSY breaking can be elegantly reached compactifying theories with
nontrivial boundary conditions (the Scherk-Schwarz mechanism). In the
framework of SW theory this leads to a possibility, for example, to formulate
the whole family of the "adjoint matter" SW theories as various limits of a
unique integrable system -- the elliptic Ruijsenaars-Schneider model
\cite{BMMM1,BMMM2}.

Consider now the compactification of \N2 SUSY Yang-Mills theory with the
only vector supermultiplet to 3+1 (compact with radius $R_3\equiv R$)
dimensions. If one takes all
fields to have the {\em periodic} boundary conditions in compact direction
this would be a \4N (in 3D sense) SUSY theory.
If, however, one puts \cite{BMMM1,BMMM2}
\be
\phi (x+R) = e^{i\epsilon}\phi (x)
\label{phase}
\ee
on {\em half} of the fields, the resulting theory would have only \N2
{\em three}-dimensional SUSY (i.e. \1N in 4D sense), i.e. the supersymmetry
will be (partially) broken by non-periodic boundary conditions.

As an example, let us consider the case of \N2 4D vector supermultiplet in
the adjoint representation, consisting of $(A_\mu,\psi)$ -- 4D \1N vector
multiplet and $(\phi,\chi)$ -- 4D \1N scalar multiplet, where $\psi$ and
$\chi$ are two complex Weyl spinors. Let the latter one acquire a nontrivial
phase (\ref{phase}) under shift along the loop in compact direction, then
it becomes massive with mass ${\epsilon\over R}$. The 4D \1N vector
multiplet remains to be massless, and can be represented as a 3D \N2
supermultiplet $(A_\alpha,\psi,{q\over R})$, where $\alpha=0,1,2$, $q=RA_3$
and $\psi $ is 3D complex spinor.

In contrast to \4N SUSY in 3D, \N2 supersymmetric theory can generate a
superpotential \cite{AHW}. Introducing complexified variables
$q_i \to q_j + i\gamma_i$ where $q_i$ are the (properly
normalized) eigenvalues of the matrix $\oint A_3$ and $\gamma_i$ are 3D dual
photons $A_i = \epsilon_{ijk}\d_j\gamma_k$, the superpotential acquires the
form \cite{SW3,n23d}
\be
W \sim {1\over R}\left( \epsilon\Tr\Phi^2 +
\Lambda^2\left(\sum_{i=1}^{N-1} e^{q_{i+1}-q_i} +
e^{q_1-q_N}\right) \right)
\ee
where the first term is "4D contribution" \cite{SW1}, the second term (first
term in the brackets) has 3D origin and the last one is induced by 3+1D
instanton contributions (see, for example, recent paper \cite{HHM} and
references therein for details). All simple roots (the first term) are usual
3D instantons (BPS "mo\-no\-po\-les") giving the potential of the {\em open}
Toda chain, while the last term (the negative root) appears only in 3+1
dimensions and can be treated as a 4D instanton (or caloron) contribution.

In the weak coupling limit this term vanishes, giving rise to the open
Toda chain \cite{SW3}. This term also vanishes in literally 3D gauge theory
(when the compact dimension shrinks to zero), it is not surprising since
dimensional  reduction implies that ${R\over g_4^2} = {1\over g_3^2}$ and
$\Lambda\sim e^{-{1\over g_4^2}} = e^{-{1\over Rg_3^2}}$ so that the 3D limit
$R\to 0$ coincides (for fixed 3D coupling $g_3$) with the weak coupling limit
in 4D gauge theory. In the weak-coupling limit (in the $SL(2)$
or "Liouville" example) the superpotential is
\be
\Lambda^2e^{2q} = u - p^2 = u - \left({dq\over dt}\right)^2
\label{liu}
\ee
it follows, that
\be
dt =
{dq\over\sqrt{u-\Lambda^2e^{2q}}}\ \stackreb{X=\Lambda^2e^{2q}}{=}\ {dX\over
2X\sqrt{u-X}}\ \stackreb{(\ref{liu})}{=}\ {dp\over p^2-u}
\ee
and integration
gives
\be p = \sqrt{u}~{1 + e^{\vartheta}\over 1 - e^{\vartheta}} =
\sqrt{u}~{1 + e^{2\sqrt{u}t}\over 1 - e^{2\sqrt{u}t}} = - \sqrt{u}
\coth{\sqrt{u}t}
\ee
so that
\be
\Lambda^2e^{2q} = X = 2u~{e^{\vartheta}\over 1 - e^{\vartheta}} =
2u\sum_{n=1}^{\infty}e^{n\vartheta}
\label{instexp}
\ee
For $\vartheta\ll 1$, the whole "instanton series" $e^{2q} \sim e^{\vartheta}$
can be reduced to the classical or "bare" result. We see, indeed, that
quantization -- at least in the sense of constructing the effective actions
-- has the form of canonical transformation of some integrable system
\footnote{In the case we literally discuss now -- the Toda chain.}.

The opposite limit $R\to\infty$ corresponds to the uncompactified 4D gauge
theory and in this limit the "3D" variables $\{ q_i\}$ or $\{\vartheta_i\}$
are {\em massive}, i.e. cannot play the role of moduli. Thus the dimension of
moduli space becomes equal to the {\em half} of the dimension of the phase
space of an integrable system and computation of the effective action
requires also an integration over the "3D" variables, or the
Bogolyubov-Whitham averaging of an integrable system. This leads to arising
of the Whitham integrable system, which can be formulated in pure geometric
terms along the lines of \cite{KriW}.

Let us point out that the SUSY breaking we have considered above is different
from a direct SUSY breaking to \1N 4D theory (without compactification)
\cite{SW1} which implies the massless monopole's limit of the \N2 spectral
curve $\Lambda^N\cosh z = P_N(\lambda )$ or, in particular, that the
polynomial $P_N(\lambda )$ (\ref{polyn}) turns into
\be\label{chebyshev}
P_N(\lambda ) = \Lambda ^N\cosh z
\ \ \ \ \ \ \ \
\lambda = 2\cosh {z\over N} \equiv \xi + \xi ^{-1}
\ee
This degeneration corresponds to the solitonic
limit of corresponding finite-gap integrable system -- for pure gauge theory
of the periodic Toda chain. In particular the curve (\ref{chebyshev}) is
a "solitonic" curve in Toda chain with the roots of polynomial $Q(\lambda )$
given by
\be
Q(\lambda ) = \prod _{j=1}^{N-1}(\lambda - 2\cos{\pi j\over N})
\ee
The generic form of the Toda BA function is
\be
\Psi_n ^{(\pm)}(\xi,t) = \xi^ne^{\sum_k t_k(\xi^k - \xi^{-k})}
{R_n ^{(\pm)}(\xi,t)\over R(\xi)}
\label{basol}
\ee
where
\be
R(\xi ) = \prod _{s=1}^{N-1}(\xi - \gamma_s)
\\
R_n ^{(\pm)}(\xi,t)=\psi_n ^{(\pm)}(t)\prod _{s=1}^{N-1}(\xi - \mu_s(n,t)) =
\sum_{k=0}^{N-1}r_k(n,t)\xi^k
\ee
and the Toda chain Lax equation
\be
\lambda\Psi_n = C_{n+1}\Psi_{n+1} + p_n\Psi_n + C_n\Psi_{n-1}
\\
C_n \equiv e^{\2(q_n-q_{n-1})}
\ \ \ \ \ \
\lambda = \xi + {1\over\xi}
\ee
implies that
\be
r_0(n) - C_nr_0(n-1) = 0
\\
r_1(n) - C_nr_1(n-1) - p_nr_0(n) = 0
\ee
i.e.
\be
r_0(n) = C_nr_0(n-1) = \dots = e^{\2(q_n-q_0)}r_0(0) \sim e^{\2q_n}
\ee
For the solitons coming from degeneration of $N$-periodic Toda chain
one should impose the "gluing conditions"
\be
\Psi_n (\xi_j) = \Psi_n({1\over\xi_j})\ \ \ \ \ \ \ j=1,\dots,N-1
\label{glue}
\ee
which mean that the BA function remembers that it came originally from
genus $N-1$ Riemann surface and each pair of points $\xi_j, {1\over\xi_j}$
corresponds to a degenerate handle. The condition (\ref{glue}) together
with the explicit form (\ref{basol}) and $\Psi_{n+N} = w\Psi_n$ gives
\be
w = \xi^N
\ \ \ \ \ \
\xi_j^{2N} = 1
\ee
i.e.
\be
\xi_j = e^{i\pi j\over N}
\ee
where the label $j$ can be restricted to $j=1,\dots,N-1$ since
\be
\phi_j = \xi_j + {1\over\xi_j} = 2\cos{\pi j\over N} = \phi_{2N-j}
\label{vacua}
\ee
Eq.~(\ref{glue}) explicitly reads
\be
{R_n({1\over\xi_j})\over R_n(\xi_j)} =
\prod_{k=1}^{N-1}{\xi_j^{-1}-\mu_k(n,t)\over\xi_j - \mu_k(n,t)} =
e^{{2\pi inj\over N}+4i\sum_l t_l \sin{\pi jl\over N}+ Z_j(\gamma)}
\\
Z_j(\gamma) \equiv \sum_{s=1}^{N-1}
\log{\xi_j-\gamma_s\over 1-\xi_j\gamma_s}
\\
j=1,\dots,N-1
\label{glue2}
\ee
This is a system of linear equations for the coefficients $r_k(n,t)$ of
the polynomial $R_n(\xi,t)$
\be
e^{i\Phi_j(n,t)\over 2}\sum_{k=1}^{N-1}
\sin\left({\pi jk\over N}+{\Phi_j(n,t)\over 2}\right) r_k = 0
\\
\Phi_j(n,t)\equiv {2\pi nj\over N}+4\sum_l t_l \sin{\pi jl\over N}-iZ_j(\gamma)
\ee
which can be easily solved.
The conditions (\ref{vacua}) can be interpreted as values of the
scalar fields in the critical points of the superpotential while the soliton
trajectories connect the critical points.

Another way to see that SUSY breaking down to \1N should correspond to the
solitonic limit comes, possibly, from more detailed study of the Whitham
hierarchy. Generating superpotential in the \1N theory can be
thought of as switching on Whitham dynamics $\delta t_k\Tr\Phi^k$
in the sense of \cite{rgwhi}.
For small values of $\delta t_k$ this can be thought as perturbation
of the smooth \N2 solution, or to the
computation of certain correlators in \N2 SYM theory. \1N theory itself
rather corresponds to finding the solutions to Whitham equations for
{\em large} values of $t_k$. The large $t_k$ asymptotic of the Whitham
solutions brings us to the boundaries of moduli space or, in other words,
corresponds to the decoupling of the smooth finite-gap solutions into
solitons.

\section{T-duality and dualities in integrable systems}

T-duality is one of the basic features of string theory in target-space with
compact dimensions and an example of coordinate-momentum duality. It is
well-known and, in fact, easy to see that the
spectrum of string on a circle of radius $R$ is invariant under the
transformation $R\to{\alpha'\over R}$ which interchanges the KK
momenta, propagating along the compact direction, with the windings of
strings along the circle. T-duality transformation also replaces the Wilson
of gauge fields loops by positions of branes or VEV's of scalars and vice
versa (see, for example, \cite{polch} for details).

Two different "pictures" of moduli space (of compactified theory) -- in terms
of Wilson lines and VEV's of scalar fields are "T-dual" to each other. This
coordinate-momentum duality can be trivially seen in the theory with \4N
SUSY when it literally corresponds to the exchange of independent parameters
$\{\phi_i\}$ and $\{q_i\}$ -- the momenta (action variables) and
co-ordinates (angles) of a trivial integrable system -- free motion of
particles on some torus. Breaking \4N SUSY leads to relating of corresponding
gauge theory with already {\em nontrivial} integrable system, when the
action variables are no longer identified with momenta. However, certain
finite-dimensional integrable systems (of the Calogero-Moser-Ruijsenaars
family) still possess nice duality properties \cite{Fock,GNR}, and it is not
quite a coincidence that this duality can be easily constructed in the systems
corresponding to the theories with KK excitations \cite{Ne96,BMMM2}.

In other words T-duality can be considered as symmetry between the $A$ and
$\Phi$ variables in the relation (\ref{momap}). In terms of integrable
systems this leads to the symmetry between two different sets of commuting
variables on the full phase space -- the original co-ordinates and
Hamiltonians (or action variables) of an integrable system \cite{Fock,GNR}.

As an example, consider, first, 2-particle trigonometric Ruijsenaars system
with the Hamiltonian
\be
h = h(p,q) = \cosh p\sqrt{1 - { m^2\over\sinh^2q}}
\ee
From the Hamiltonian equations
\be
{dq\over dt} = {\d h\over\d p} = \sinh p\sqrt{1 - { m^2\over\sinh^2q}}
\\
{dp\over dt} = - {\d h\over\d q} = { m^2\cosh p\cosh q\over
\sinh^3 q\sqrt{1 - { m^2\over\sinh^2q}}}
\ee
it follows that
\footnote{In this form the equation of motion coincides exactly with the {\em
non}-relativistic limit -- the trigonometric Calogero model, the same effect as
in relativistic and non-relativistic Toda chains (see, for example \cite{FM}).}
\be
{d^2 q\over dt^2} = {\d\over\d q}\left({ m^2\over 2\sinh^2 q}\right)
\ee
or
\be
\left({dq\over dt}\right)^2 - { m^2\over\sinh^2 q} = E
\label{caleq}
\ee
with $E = h^2 -1$. Solving (\ref{caleq}) one gets
\be
\sqrt{E}\ t = \log\left(\cosh q + \sqrt{{ m^2\over E} + \sinh^2 q }\right)
- \log\sqrt{1 - { m^2\over E}}
\label{solsl2}
\ee
(with a particular choice for the integration constant).
It is easy to check that the symplectic form is
\be
\Omega = dp\wedge dq = dh\wedge dt = dx\wedge d\pi
\ee
where $h = \cosh x$, $\pi = t\sqrt{E} = t\sqrt{h^2-1}$. Moreover it is easy to
see that, introducing $H = \cosh q$, one gets from Eq.~(\ref{solsl2})
\be
H = \2\sqrt{1 - { m^2\over E}}\left( e^{\pi} + e^{-\pi}\right) =
\cosh\pi\sqrt{1 - { m^2\over\sinh^2x}}
\ee
i.e. the Hamiltonian of the {\em dual} system which is again the trigonometric
Ruijsennars model with the {\em same} coupling constant $m^2$.

This {\em self}-duality of trigonometric Ruijsenaars
system turns into duality between the trigonometric Calogero-Moser model
and rational relativistic Ruijsenaars model in almost obvious
"nonrelativistic" limit. The equation of motion
(\ref{caleq}) can be equally considered as an equation of motion for
the non-relativistic Calogero-Moser model with the Hamiltonian
\be
h_{CM} = \2 p^2 - { m^2\over\sinh^2q}
\label{trical}
\ee
so that $h_{CM} = E = \sqrt{h^2-1}$. The result of integration of the
equation of motion is again (\ref{solsl2}) but now one has to consider
rather "non-relativistic" limit of "small" $x$ and "p", i.e.
$h_{CM}= E = x^2$ but still $\cosh q = H$. It follows then, that the
system (\ref{trical}) is dual to rational Ruijsenaars model with the
Hamiltonian
\footnote{
Going further, it is easy to check that in the "double" non-relativistic
limit one comes to the (again self-dual) rational Calogero model with
\be
h_C = p^2 - { m^2\over q^2} = x^2
\ee
or
\be
H_C = \pi^2 - { m^2\over x^2} = q^2
\ee}
\be
H = \cosh\pi\sqrt{1-{ m^2\over x^2}}
\ee
In general $N$-particle trigonometric Ruijsenaars-Schneider system it was
shown in \cite{Fock} that the duality transformation can be interpreted as
modular transformation in the
space of the $SL(N)$ valued flat connections on torus. These flat
connections are described by two $SL(N)$ matrices in general position, say,
$(A,B)$ modulo common conjugation: $(A,B) \mapsto (gAg^{-1},gBg^{-1})$.
According to \cite{Fock}, this space is endowed with the Poisson bracket
\be
\{ A \stackreb{,}{\otimes} A\} = r_a A\otimes A + A\otimes A r_a +
(1 \otimes A) r_{21}(A \otimes 1) - (A \otimes 1) r_{12} (1 \otimes A)
\\
\{ A \stackreb{,}{\otimes} B\} = r_{12} A\otimes B + A\otimes B r_{12} +
(1 \otimes B) r_{21} (A \otimes 1) - (A \otimes 1) r_{12} (1 \otimes B)
\\
\{ B \stackreb{,}{\otimes} B\} = r_a B\otimes B + B\otimes B r_a +
(1 \otimes B) r _{21}(B \otimes 1) - (B \otimes 1) r_{12} (1 \otimes B)
\label{pbfr}
\ee
with
\be
r_{12} = \sum_{\alpha > 0} E_{\alpha} \otimes E_{-\alpha} + \frac{1}{2}
\sum_i H_i \otimes H_i
\\
r_{21} = \sum_{\alpha > 0} E_{-\alpha} \otimes E_{\alpha} + \frac{1}{2}
\sum_i H_i \otimes H_i
\\
r^a = \frac{1}{2}\left( r_{12} - r_{21}\right)
\ee
which is degenerate, but can be inverted, for example, on a symplectic
leave, defined as
\be
ABA^{-1}B^{-1} =  m^2 {\bf 1} + R^{(1)}
\label{list}
\ee
where $R^{(1)}\equiv\bxi\otimes\bbeta $ is a matrix of unit rank
(or $R^{(1)}_{ij}\equiv\xi_i\eta_j $ with some, dependent on $A$ and $B$,
vectors $\bxi $ and $\bbeta $). Diagonalizing, for example,
$A = {\rm diag}(Q_1,\dots,Q_N)$ one can turn the relation (\ref{list})
into
\be
\left( {Q_i\over Q_j} -  m^2\right)B_{ij} = \tilde{R}^{(1)}_{ij} =
\xi_i\tilde\eta_j
\ee
where $\tilde\bbeta = \bbeta\cdot B$. Now, using the freedome of conjugation
by diagonal matrices (which leaves the diagonal form of $A$ untouched) one can
put (in general position) all $\xi_j = 1$, ending up with
\be
B_{ij} = {\tilde{\eta}_iQ_i \over Q_i -  m^2 Q_j}
\ee
The traces of this matrix, as functions of $\tilde{\eta}_i=e^{p_i}$ and
$Q_i=e^{q_i}$ give Hamiltonians of the trigonometric Ruijsenaars-Schneider
 system. In the nonrelativistic limit equation (\ref{list}) turns into
the commutator relation
\be [A,B] = m^2 {\bf 1} + R^{(1)}
\label{momap1}
\ee
with the solution
\be
B_{ij} = p_i\delta_{ij} + { m^2\over Q_i-Q_j}
\ee
As we discussed in sect.~\ref{s:pert} the trigonometric Ruijsenaars-Schneider
system literally corresponds to the perturbative limit of the SW theory with
the adjoint mass and KK excitations (\ref{spectrum}), so this is the way how
perturbative T-duality transformation is realized on the phase space of
corresponding integrable system.

\section{Conclusion}

In these notes we have tried to review main ingredients of the approach based
on relation of the Seiberg-Witten effective theories with integrable systems.
Recent studies have shown that the existence of the exact
non-perturbative integrable differential equations allows to compute
explicitly some physical quantites in 4D SUSY gauge theories.

Moreover, it turns out that an integrable system in a most straightforward
way is seen in the compactified SUSY gauge theory -- in 3 plus 1 compact
dimensions. In this case, one finds that the symplectic transformation is
nothing but a change of variables from bare to exact quantum variables and
the set of Hamiltonians (or the spectral curve equation) arises as
superpotentials in the compactified theory with broken SUSY.

{\bf Note added}. In a very recent paper \cite{Dorey} one can find very
similiar conclusions concerning the relation between superpotentials and
structures of the integrable system we have discussed above. Moreover,
this paper contains the detailed analysis of breaking SUSY down to \1N in the
theory with finite adjoint mass -- the corresponding superpotential
coincides, like it should follow from general reasoning and the results of
\cite{SWCal}, with that of elliptic Calogero-Moser model.

\section*{Acknowledgements}

I am indebted to H.Braden, E.Corrigan, V.Fock, A.Gerasimov, S.Kharchev,
A.Losev, A.Mironov, A.Morozov, A.Rosly and B.Voronov for illuminating
discussions and
I am grateful to T.Inami, R.Sasaki, T.Uematsu, all other organizers of the
conference in Kyoto and T.Takebe for warm hospitality in Japan. The work was
also supported by the RFBR grant 98-01-00344 and the INTAS grant 99-0103.

\end{document}